\documentclass[aps,prb,reprint,showpacs,longbibliography,superscriptaddress]{revtex4-1}
\usepackage{bm}
\usepackage{graphicx}
\usepackage{epstopdf}
\usepackage{latexsym}
\usepackage{subfigure}
\usepackage{color}
\usepackage{hyperref}
\usepackage{amssymb}
\usepackage{amsmath}

\begin{document}

\title{Spin gap in a quasi-1D $\bf S = \frac{1}{2}$ antiferromagnet K$_2$CuSO$_4$Cl$_2$}
\author{T.~A.~Soldatov}
\affiliation{P.~L.~Kapitza Institute for Physical Problems, RAS, 119334 Moscow, Russia} \affiliation{Moscow Institute for Physics and Technology,
141701 Dolgoprudnyi, Russia}

\author{A.~I.~Smirnov}
\affiliation{P.~L.~Kapitza Institute for Physical Problems, RAS, 119334 Moscow, Russia}

\author{{K.~Yu.~Povarov}}
\affiliation{Laboratory for Solid State Physics, ETH Z\"{u}rich, 8093 Z\"{u}rich, Switzerland}

\author{M. H\"{a}lg}
\affiliation{Laboratory for Solid State Physics, ETH Z\"{u}rich, 8093 Z\"{u}rich, Switzerland}

\author{W. E. A. Lorenz}
\affiliation{Laboratory for Solid State Physics, ETH Z\"{u}rich, 8093 Z\"{u}rich, Switzerland}

\author{A. Zheludev}
\affiliation{Laboratory for Solid State Physics, ETH Z\"{u}rich, 8093 Z\"{u}rich, Switzerland}

\begin{abstract}

Electron spin resonance experiments in the quasi-1D $S = \frac{1}{2}$  antiferromagnet K$_2$CuSO$_4$Cl$_2$ reveal opening of a gap in absence of
magnetic ordering, as well as an anisotropic shift of the resonance magnetic field. These features of magnetic excitation spectrum are explained
by a crossover between a gapped spinon-type doublet ESR formed in a 1D antiferromagnet with uniform Dzyaloshinskii-Moriya interaction and a
Larmor-type resonance of a  quasi-1D Heisenberg system.

\end{abstract}

\date{\today}

\maketitle

\section{Introduction}\label{Introduction}

Heisenberg  antiferromagnetic spin $S = \frac{1}{2}$ chain has a critical ground state, disordered because of quantum fluctuations. In a  3D
crystal, containing  spin chains coupled due to  a weak interchain exchange, magnetic ordering occurs at a N\'{e}el temperature $T_{N}$ far below
the Curie-Weiss temperature $T_{CW}$. Strong spin-spin correlations appear below $T_{CW}$, therefore, in a temperature range $T_{N} < T \ll
T_{CW}$, the spin structure and excitations of a quasi-1D $S = \frac{1}{2}$ antiferromagnet are close to that of a quantum spin-liquid of
noninteracting spin chains at $T=0$.  The characteristic feature of this quantum spin liquid is a gapless continuum of $\Delta S$=1 transitions.
These transitions correspond to creation of pairs of fractional excitations carrying spin $S = \frac{1}{2}$ (called "spinons"), see, {\it e.g.}
Refs.\cite{Faddeev,Tennant1,Tennant2} The continuum was intensively studied by neutron scattering at wavevectors of the order of the reciprocal
lattice period, where the width of the continuum is of the order of maximum energy. At small wavevectors near the center of the Brillouin zone the
width of the continuum is negligible and the spectrum is mainly analogous to that of conventional spin waves. However, in presence of  the uniform
Dzyaloshinskii-Moriya (DM) interaction of magnetic ions within a chain, there arises a remarkable transformation of the two-spinon continuum at
$q\rightarrow 0$, causing a spin gap and a shift of a magnetic resonance. \cite{Gangadharaiah,Povarov,Karimi} This transformation may be detected
in electron spin resonance (ESR) experiments.

 DM interaction between two magnetic ions with the spins ${\bf
S}_i$ and ${\bf S}_j$ has a contribution to the Hamiltonian ${\bf D}_{ij}\cdot{\bf S}_i  \times {\bf S}_j$. In most of antiferromagnetic crystals
with DM interaction, the characteristic vectors ${\bf D}_{ij}$ form a staggered structure, which causes a canting  of sublattices in a
conventional  antiferromagnet. \cite{Dzyaloshinsky,Moriya} For the {\it uniform} DM interaction vectors ${\bf D}_{ij}$ are parallel. This results
in a  spiral spin structure with a wavevector $q_{DM} = D/(Ja)$ ($a$ is the lattice constant) for a classical antiferromagnet.

In a quantum 1D antiferromagnet there is no ordering in a ground state, however, the uniform DM interaction modifies the spectrum of a continuum
by a shift in $q$-space for the same vector $q_{DM}$. \cite{Gangadharaiah,Povarov,Karimi} As a result, a spin gap of the order of $D$ arises in
zero field and a modification of the ESR frequency occurs:  in a magnetic field parallel to $\bf D$ the ESR spectrum transforms from a single line
at a Larmor frequency to a doublet. The split of the doublet equals the width of the initial continuum at $q=q_{DM}$. These features in the ESR
spectrum were observed for the quasi 1D S=1/2 antiferromagnets  with uniform DM interaction within chains: Cs$_2$CuCl$_4$
\cite{Povarov,Smirnov1,Smirnov1a} and K$_2$CuSO$_4$Br$_2$.\cite{Halg,Smirnov2} The two examples present almost the complete list of the known
compounds with the uniform DM interaction. The third spin-chain compound with the uniform DM-interaction is K$_2$CuSO$_4$Cl$_2$. \cite{Halg} This
compound  has  much weaker intrachain exchange integral (3.2 K), than  Br-compound (20.4 K), at the same time the interchain exchange integral
$J'$ calculated for both compounds from the value of the N\'{e}el temperature is nearly the same, i.e. about 0.03 K. $T_N$  is about 100 mK for
the bromide and 77 mK for the chlorine compound, according Ref.\cite{Halg} The direct measurement of the interchain exchange by the inelastic
neutron scattering in the saturated phase results in a value of 0.45~K for K$_2$CuSO$_4$Cl$_2$, \cite{Blosser} which is about 15 times greater,
than the estimated from N\'{e}el temperature. Anyway, both these $J'$ values indicate, that the Cl-compound is less one-dimensional compared to
K$_2$CuSO$_4$Br$_2$, which has $J/J' \simeq$  600. The value of the energy of DM interaction should be diminished in the Cl-compound with respect
to Br-compound, proportional to $J$, as follows from theory. \cite{Moriya} Thus, the hierarchy of interactions in K$_2$CuSO$_4$Cl$_2$ is expected
to be as $J \gg J^\prime \gtrsim D$ instead of  $J \gg   D \gg J^\prime$ in case of K$_2$CuSO$_4$Br$_2$.  This provides a possibility to look here
in K$_2$CuSO$_4$Cl$_2$ for the spinon effects in presence  of interchain exchange competing with uniform DM interaction and compare the
observations with the almost ideal 1D case of K$_2$CuSO$_4$Br$_2$.  The differences between $J \gg J^\prime \gg D$ and $J \gg   D \gg J^\prime$
situations are predicted to  lead to very different phase diagrams for corresponding materials in the ordered, low-temperature limit.
\cite{JinStarykh} Therefore the determination of the relation between $J^\prime$  and  $D$ is of importance for the spin structure and phase
transitions of low-temperature ordered phases.

\begin{figure}[b]
\begin{center}
\includegraphics[width=0.42\textwidth]{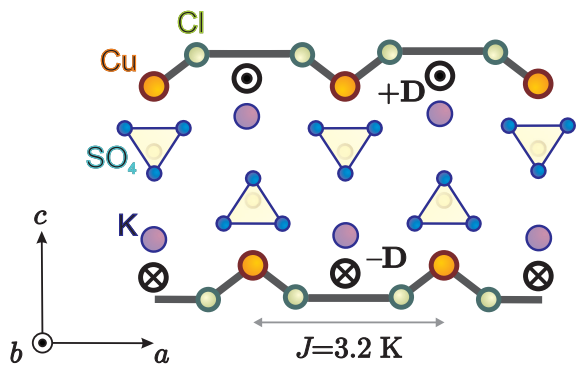}
 \caption{\label{fig1} Sketch of crystallographic structure of K$_2$CuSO$_4$Cl$_2$.}
\end{center}
\end{figure}

 The crystal structure for both K$_2$CuSO$_4$Br$_2$ and
K$_2$CuSO$_4$Cl$_2$ is  orthorombic  with space group $Pnma$. The magnetic $S = \frac{1}{2}$ Cu$^{2+}$ ions are linked by the superexchange via
two halogen atoms in chains running along the $a$ axis.  The symmetry analysis of Ref.\cite{Halg} shows that DM vectors $\bf D$ are directed
parallel to $b$ axis and pointed in the same direction within a chain but antiparallel in adjacent chains, as shown in Fig.~\ref{fig1}.

\begin{figure}[t]
\begin{center}
 \includegraphics[width=0.5\textwidth]{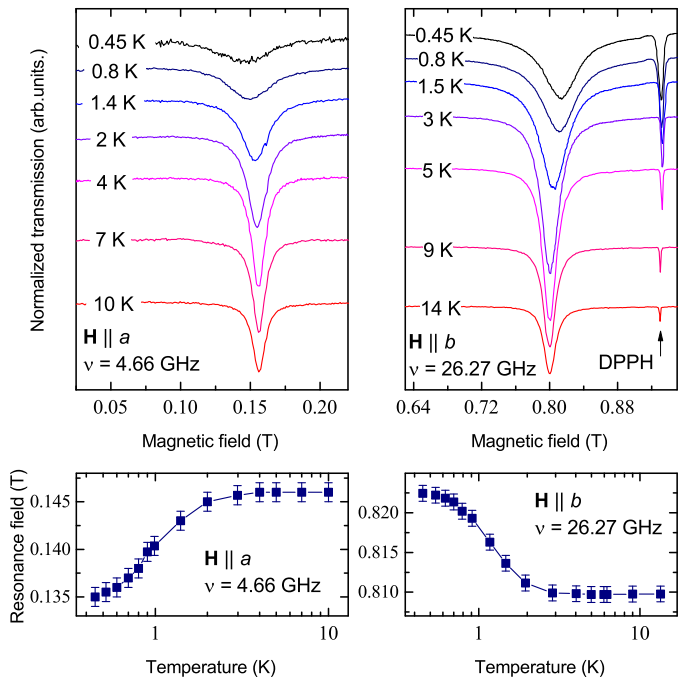}
 \caption{\label{fig2} (Color online)
Temperature evolution of ESR line (upper panels) and temperature dependence of ESR field (lower panels)
 for ${\bf H }\parallel a$ at $\nu$~=~4.66~GHz (left panels) and for ${\bf H }\parallel b$ at $\nu$~=~26.27~GHz (right panels).
 Transmission is normalized to have a linear dependence on the imaginary part of susceptibility (see text).
 The records of transmission are vertically offset for clarity.}
\end{center}
\end{figure}

Hamiltonian of an isolated spin chain with uniform DM interaction may be written as follows:
\begin{eqnarray}
\mathcal{H} = \sum_{i} (J{\bf S}_{i}\cdot{\bf S}_{i+1} + {\bf D}\cdot{\bf S}_{i}\times{\bf S}_{i+1} + g\mu_{B}{\bf H}\cdot{\bf S}_{i}) \ \
\label{Ham}
\end{eqnarray}

From here, in a low-frequency approximation ($2\pi \hbar \nu \ll J$), the $T$=0 ESR frequencies were derived  for the arbitrary direction of the
magnetic field $\bf H$: \cite{Gangadharaiah,Karimi,Povarov}
\begin{eqnarray}
2\pi \hbar \nu = \left | g\mu_{B}{\bf H} \pm \frac{\pi}{2}{\bf D} \right |.
\label{Freq}
\end{eqnarray}

For ${\bf H} \parallel {\bf D}$ this results in a doublet of ESR lines instead of a single line at $D=$0. Besides, a gap of $\pi D/2$ is predicted
for zero field.

\begin{figure}[b]
\begin{center}
\includegraphics[width=0.5\textwidth]{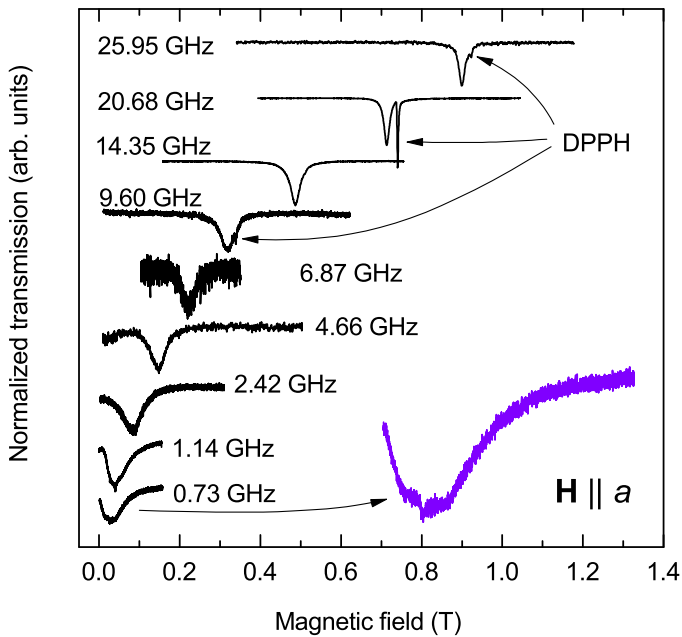}
 \caption{\label{fig3} (Color online) Examples of normalized ESR records taken at $T=$ 0.45~K for ${\bf H }\parallel a$. The zoom of 0.73 GHz ESR line is 4-fold for
 vertical and horizontal scale.}
\end{center}
\end{figure}

The goal of the present work is to probe the ESR spectrum when the interchain exchange competes with DM interaction in a formation of the low
energy  dynamics of a quasi-1D antiferromagnet. We observe a strong anisotropy of the low-temperature deviation of the magnetic resonance
frequency in agreement  with (\ref{Freq}) and  the formation of a predicted zero-field energy gap. These observations indicate a peculiar
transformation of the continuum in a 1D chain with uniform DM interaction. However, we do not observe the upper component of the spinon doublet.
This reveals a crossover between a gapped spinon-type doublet ESR formed in a 1D antiferromagnet with uniform DM interaction and a Larmor-type
resonance of a Heisenberg system. Experiment shows, that the doublet is not finally formed, but the resonance is already shifted from the Larmor
frequency. To check the qualitative correspondence of the ESR spectra to spinon-type ESR  with a gap $\pi D/2$, we alternatively estimate the
value of $D$ from the high-temperature ESR linewidth, using a theory of Ref. \cite{Fayzullin} As a result, we get a qualitative agreement, natural
for the essential deviations from ideal 1D model with $J^\prime = 0$.

\section{Experimental details}\label{Details}

The magnetic resonance signals were recorded at a fixed frequency as field dependencies of the microwave power transmitted through the resonator.
For weak absorption in the sample, the change of the transmitted power is proportional to the imaginary part of the magnetic susceptibility
$\chi^{\prime\prime}$.

 For a strong absorption, e.g. for the narrow ESR in the paramagnetic region, the linear correspondence between the change of the
transmission  and $\chi^{\prime\prime}$ may be lost. In case when microwave losses in the sample are comparable or exceed that in the walls of the
resonator, the signal just drops almost to zero. At further growth of $\chi^{\prime\prime}$ the absolute changes of the transmission are small. To
reconstruct the effective signal, which  diminishes proportional to $\chi^{\prime\prime}$, a normalization of the transmitted microwave signal $u$
was made. We use the relation between the loaded cavity $Q$-factor $Q_l$ and the losses in the sample: $1/Q_l=1/Q_0+\alpha \chi^{\prime\prime}
V_s$, here $Q_0$ is the loaded $Q$-factor of the resonator without the sample, i.e. in a magnetic field far from resonance, $\alpha$ is a
geometric factor, and $V_s$ is the sample volume. The analysis of the transmission (following, e.g. Ref.\cite{Pool}) gives for the effective
signal $\tilde{u}$:

\begin{eqnarray}
\tilde{u} = u_{0} ( 2-\sqrt{\frac{u_{0}}{u}} ) \ \ \label{largesignal}
\end{eqnarray}

  The ESR absorption lines were
recorded using a set of microwave spectrometers equipped with superconducting 12~T solenoid and $^{3}$He pump refrigerator providing temperature
down to 0.45~K. The sample was mounted inside the copper resonator which was placed in vacuum and connected via a heat link to the $^{3}$He
chamber. Angular dependencies of ESR line were taken in a spectrometer equipped with 8~T solenoid and $^{4}$He pump refrigerator with the lowest
temperature of 1.3~K.

\begin{figure}[t]
\begin{center}
\includegraphics[width=0.5\textwidth]{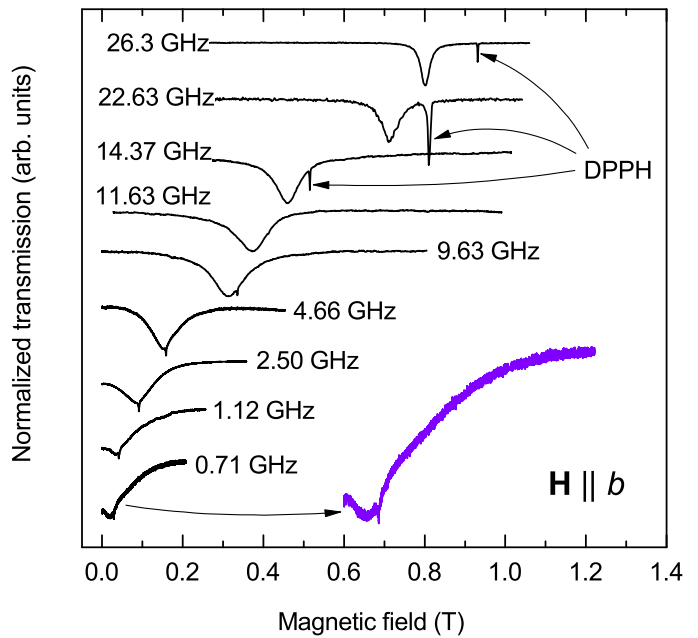}
 \caption{\label{fig4} (Color online) Examples of normalized ESR records taken at $T=$0.45~K for ${\bf H }\parallel b$. The zoom of 0.71 GHz ESR line is 3-fold for vertical and horizontal scale.}
\end{center}
\end{figure}

\begin{figure}[t]
\begin{center}
\includegraphics[width=\columnwidth]{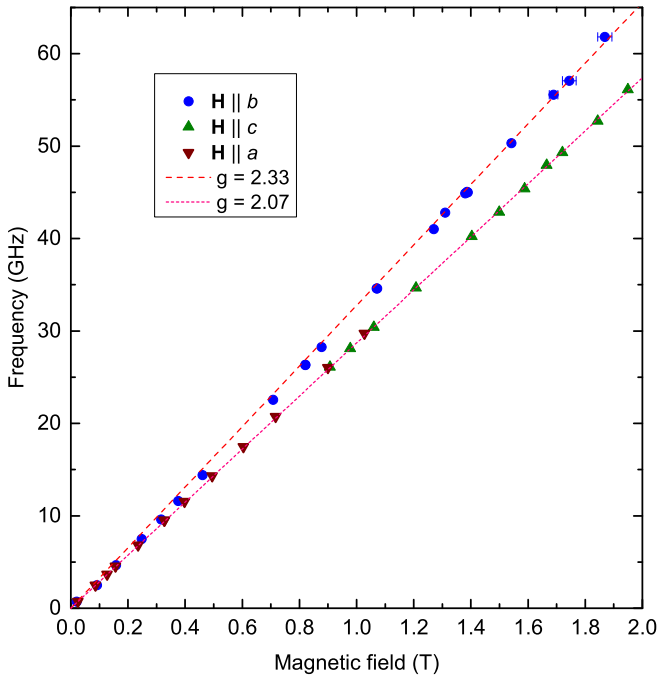}
 \caption{\label{freq-field-diagr} (Color online) Frequency-field diagram for  ESR at $T=$0.45 K
for ${\bf H }\parallel a$,  ${\bf H }\parallel b$, ${\bf H }\parallel c$.}
\end{center}
\end{figure}

\begin{figure}[t]
\begin{center}
\includegraphics[width=\columnwidth]{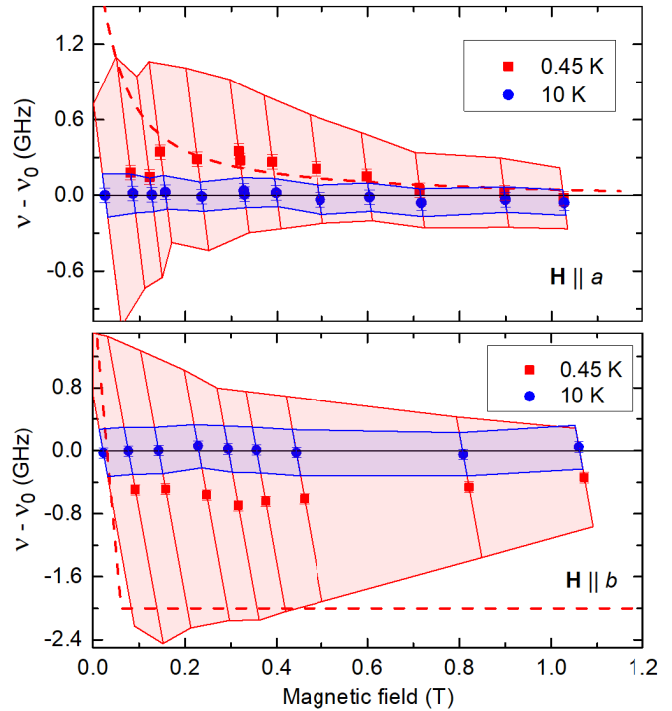}
 \caption{\label{fig5} (Color online) Shift of the resonance frequencies relative to paramagnetic resonance for ${\bf H }\parallel a$
 (upper panel) and ${\bf H }\parallel b$ (lower panel). Dashed lines are theoretical dependencies for gapped mode and lower frequency
 component of spinon doublet according to (\ref{freqPERP}) and (\ref{freqPAR}) with $\Delta = \pi D/2$~=~2~GHz. Shaded areas present
 the range of absorption exceeding a half of the maximum absorption value.}
\end{center}
\end{figure}

\begin{figure}[t]
\begin{center}
\vspace{-0.5cm}
\includegraphics[width=0.5\textwidth]{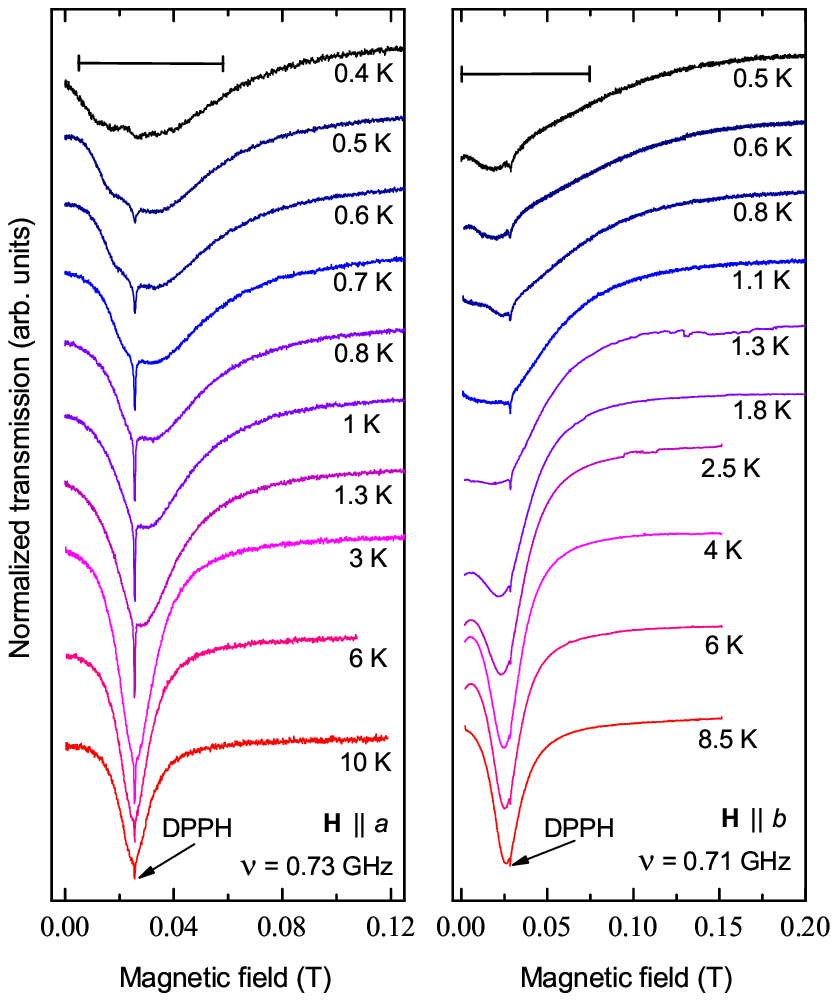}
 \caption{\label{fig8} (Color online) Temperature evolution of ESR line for ${\bf H }\parallel a$ at $\nu$~=~0.73~GHz (left panel),
  for ${\bf H }\parallel b$ at $\nu$~=~0.71~GHz (right panel). Solid horizontal line segment indicates area of ESR absorbtion exceeding 50~$\%$
  of its maximum magnitude for ESR line taken at the lowest temperature. Transmission is normalized and transformed to have a
  linear dependence on the imaginary part of susceptibility (see text). The records of transmission are
vertically offset for clarity.}
\end{center}
\end{figure}

Experiments were performed in the frequency range 0.5~-~60~GHz using Gunn diodes and klystrons as microwave sources. The microwave units with
cylindrical, rectangular, cut-ring, and spiral resonators were used for recording the resonance absorption of microwaves. A RuO$_{2}$ thermometer
monitored the temperature of the resonator. The temperature of the sample was controlled by use of a heater placed on the microwave unit.
2,2-diphenyl-1-picrylhydrazyl (DPPH) was used as a standard g = 2.00 marker for the magnetic field.

We used single crystals of K$_2$CuSO$_4$Cl$_2$ with the mass 5-40 mg from the same batch as in Ref. \cite{Halg}

\section{Experimental results}\label{Results}

At temperatures above 3~K we observe a single  Lorentzian  ESR absorption line. At these temperatures, in the whole frequency range, the resonance
field corresponds to a temperature independent $g$-tensor with principal axes coinciding with crystallographic axes. The  principal values  of
$g$-tensor are $g_{a} = g_{c} = g_{\perp} = 2.07 \ \pm \ 0.01$, $g_{b} = 2.33 \ \pm  \ 0.01$.

The temperature evolution of the 4.66~GHz ESR line at ${\bf H }\parallel a$ is shown in the left upper panel of Fig.~\ref{fig2}. Here magnetic
field is oriented perpendicular to the DM vector $\bf D$. Upon cooling to 0.45~K the ESR line is broadening and shifting to lower fields. The
corresponding dependence of the resonance field on temperature is presented in the left lower panel of Fig.~\ref{fig2}. On the contrary, at ${\bf
H }\parallel b$ (this is parallel  to  $\bf D$), the ESR line exhibits a shift to higher fields, see temperature evolution of ESR line and the
dependence of 26.27~GHz ESR field on temperature  in the right panels of Fig.~\ref{fig2}.  The shift of the ESR field at ${\bf H }\parallel c$ is
of the same value and sign as for ${\bf H }\parallel a$.

\begin{figure}[t]
\begin{center}
\vspace{-0.5cm}
\includegraphics[width=0.5\textwidth]{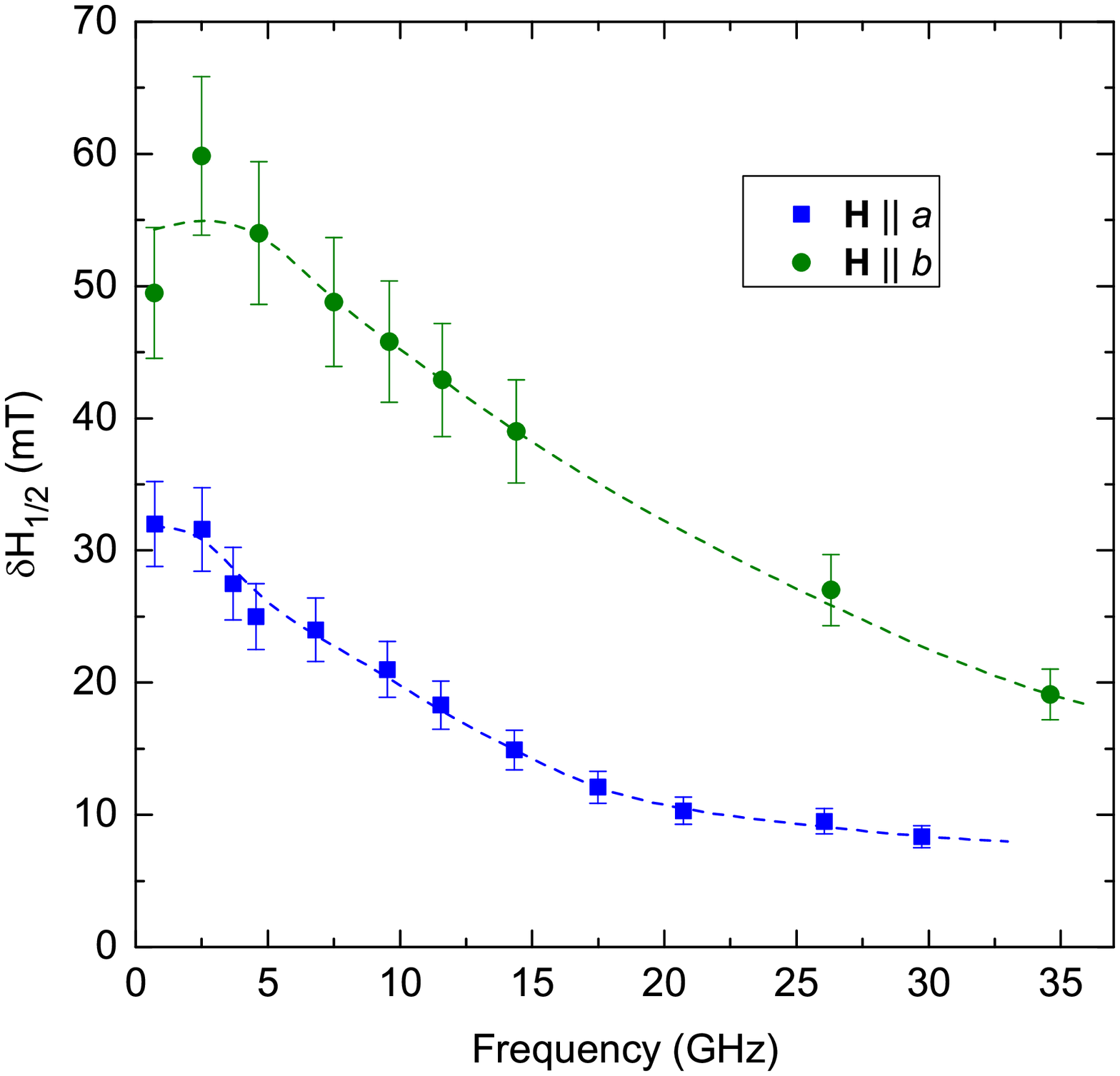}
 \caption{\label{dHvsF} (Color online) Linewidth  vs frequency dependencies of ESR  at T=0.45 K ${\bf H }\parallel a$  and ${\bf H }\parallel b$.}
\end{center}
\end{figure}

\begin{figure}[b]
\begin{center}
\includegraphics[width=0.5\textwidth]{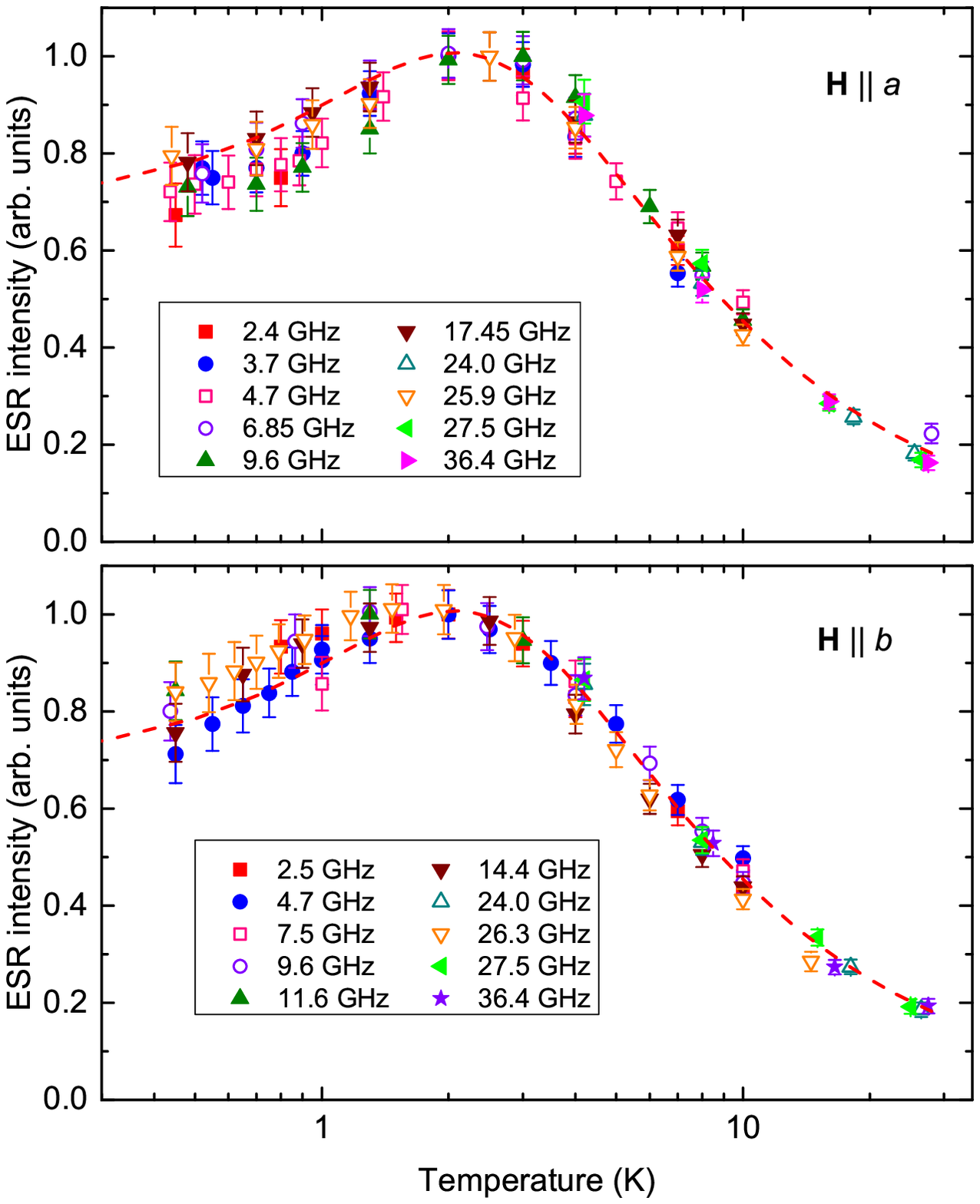}
 \caption{\label{fig6} (Color online) Temperature dependencies of ESR integral intensity at different frequencies from the range
$\nu$~=~2.4~--~36.4~GHz for ${\bf H }\parallel a$ (upper panel) and ${\bf H }\parallel b$ (lower panel). Dashed lines in both panels represent
normalized static magnetic susceptibility of $S = \frac{1}{2}$ Heisenberg chain according to Bonner-Fisher calculation for $J = 3.2$~K
\cite{Hatfield}.}
\end{center}
\end{figure}

Examples of ESR records taken at different frequencies for ${\bf H }\parallel a$ and ${\bf H }\parallel b$ at T~=~0.45~K are presented in
Figs.~\ref{fig3} and \ref{fig4} correspondingly.  The resonance fields measured at different frequencies are plotted on the  frequency-field
diagram Fig. \ref{freq-field-diagr} for three principal directions of the magnetic field. The frequency-field dependencies  mainly follow the
linear dependence observed for the high-temperature regime at $T=$10 K. The shift of the resonance field observed at cooling below 3 K is small in
comparison with the applied field, therefore the shift of the ESR frequency $\Delta \nu$ from the Larmor-type frequency $\nu_{0} = g_\alpha\mu_B
H_\alpha/(2 \pi \hbar)$ is shown separately in Fig.~\ref{fig5}. For ${\bf H }\parallel a$ the frequency shift is positive indicating opening of
the gap. This deviation monotonically increases with decreasing the field to about 0.15 T, while below this value the lineform becomes
non-Lorentzian. The resonance line transforms to a band of absorption at ${\bf H }\parallel a$ (lower record on Fig.\ref{fig3}) or to a
nonsymmetric line with a wide right wing at ${\bf H }\parallel b$ (lower record on Fig.\ref{fig4}). The maximum of absorption moves back to Larmor
frequency in fields below 0.15 T.

The transformation of the low-frequency ($\nu <$~2.4~GHz) line into a  broad band of absorbtion with an asymmetric shape is presented by the
transmission records in Fig.~\ref{fig8} both for for ${\bf H }\parallel a$ and  ${\bf H }\parallel b$. The corresponding widening of the
absorption range is illustrated in  Fig.\ref{fig5}. Here the deviation of the resonance frequency from the Larmor frequency is shown  by symbols
and the range of the magnetic field corresponding to the absorption exceeding a half of the maximum value, converted to the frequency range using
the observed $\nu(H)$ dependence is presented by segments. Resonance fields for $\nu
>$~2.4~GHz are obtained by  the fits of the ESR absorption curves  by a sum of two
Lorentzian curves with symmetric resonance fields $\pm H_{res}$. For non-Lorentzian absorption curves, occuring at $\nu <$~2~GHz the symbols
present the field of the maximum absorption and segments - the field interval where the absorption is higher then half of the maximum value.
Finally, the area of absorption exceeding a half of the maximum value at $T=$0.45~K is presented by the pink shaded area on Fig~\ref{fig5}. At the
temperature $T=10$ K we observe narrow Lorentzian resonances with a field-independent linewidth (half width at half height, $\delta
H_{\frac{1}{2}}$) of about 4 mT for ${\bf H}\parallel a $  and 8 mT for ${\bf H}
\parallel b $.
The area of $T=$ 10~K absorption exceeding a half of the maximum value is presented as shaded in blue. We see, that at low temperature of 0.45 K
there is essential frequency-dependent broadening of the ESR line, as presented quantitatively in Fig. \ref{fig5} and Fig. \ref{dHvsF}.


\begin{figure}[t]
\begin{center}
\vspace{-0.5cm}
\includegraphics[width=0.5\textwidth]{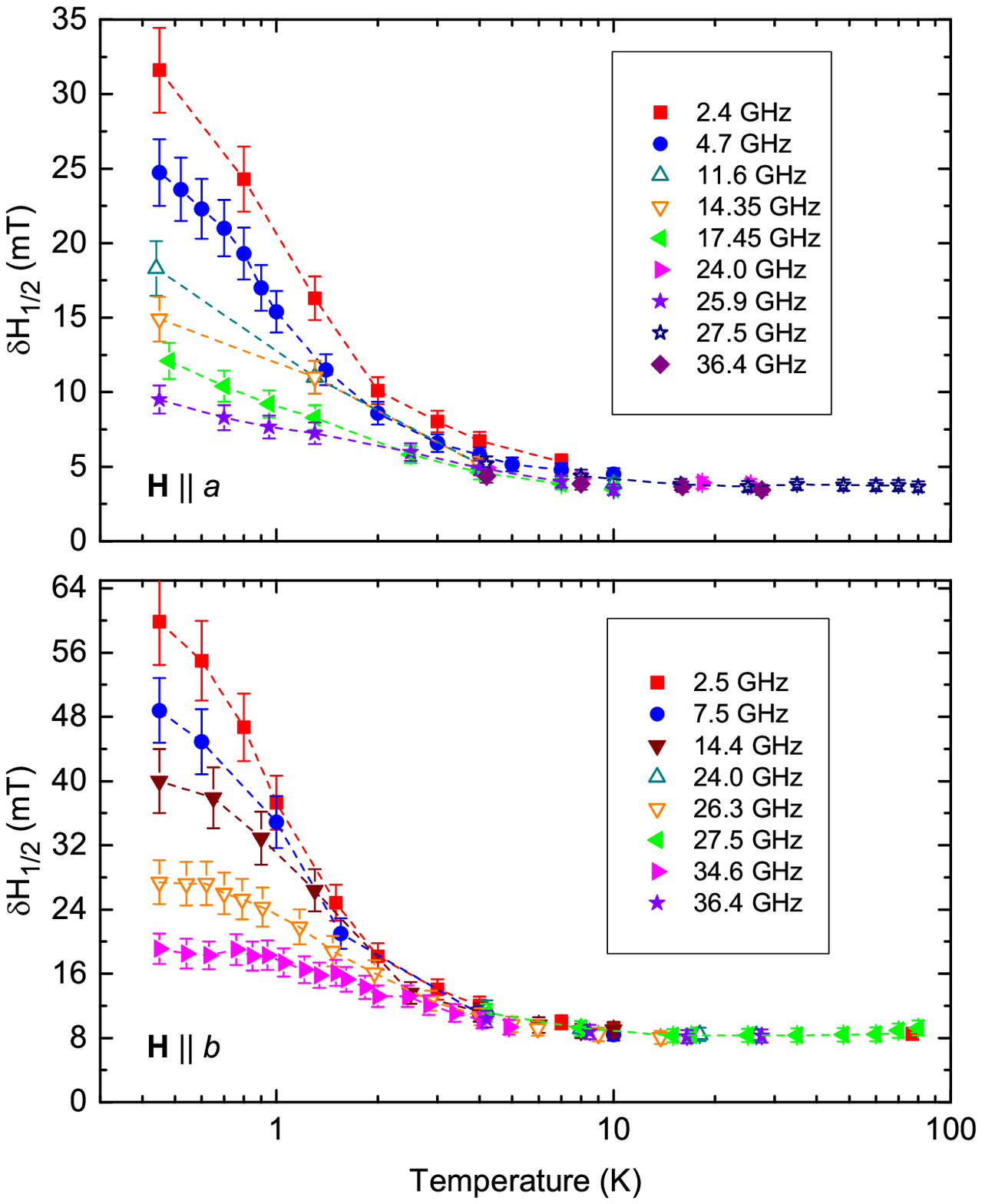}
 \caption{\label{fig7} (Color online) Temperature dependencies of ESR half width at half maximum $\delta H_{1/2}$ at different frequencies from the range $\nu$~=~2.4~--~36.4~GHz for ${\bf H }\parallel a$ (upper panel) and ${\bf H }\parallel b$ (lower panel).}
\end{center}
\end{figure}

 In the magnetic fields above 0.15 T (i.e. for frequencies $\nu >$~2.4~GHz)  ESR lines have a regular Lorentzian form. For these lines the
 integral intensity may be derived reliably. At the temperature above 3 K, both at  ${\bf H }\parallel a$ and ${\bf H
}\parallel b$, the integral intensity follows the Curie-Weiss law $\propto 1/(T + T_{CW})$ with Curie-Weiss temperature $T_{CW} \simeq 3$~K.
Integral intensity reaches its maximum value at a temperature 2~K and falls a bit at cooling below 2 K, in agreement with the static magnetic
susceptibility of $S = \frac{1}{2}$ Heisenberg chain, \cite{Hatfield} see Fig.~\ref{fig6}. The linewidth  $\delta H_{\frac{1}{2}}$, increases
monotonically on cooling below 80~K rising from 3.5~$\pm$~0.3~mT at ${\bf H }\parallel a$ and from 8.5~$\pm$~0.5~mT at ${\bf H }\parallel b$. The
dependencies of $\delta H_{\frac{1}{2}}$ on temperature taken at different frequencies  are shown in Fig.~\ref{fig7}. As mentioned above, the
low-temperature broadening of the ESR line depends strongly on frequency (see Figs.~\ref{fig3}, \ref{fig4}, \ref{dHvsF}) with a stronger
broadening at low frequencies.

\begin{figure}[t]
\begin{center}
 \includegraphics[width=0.5\textwidth]{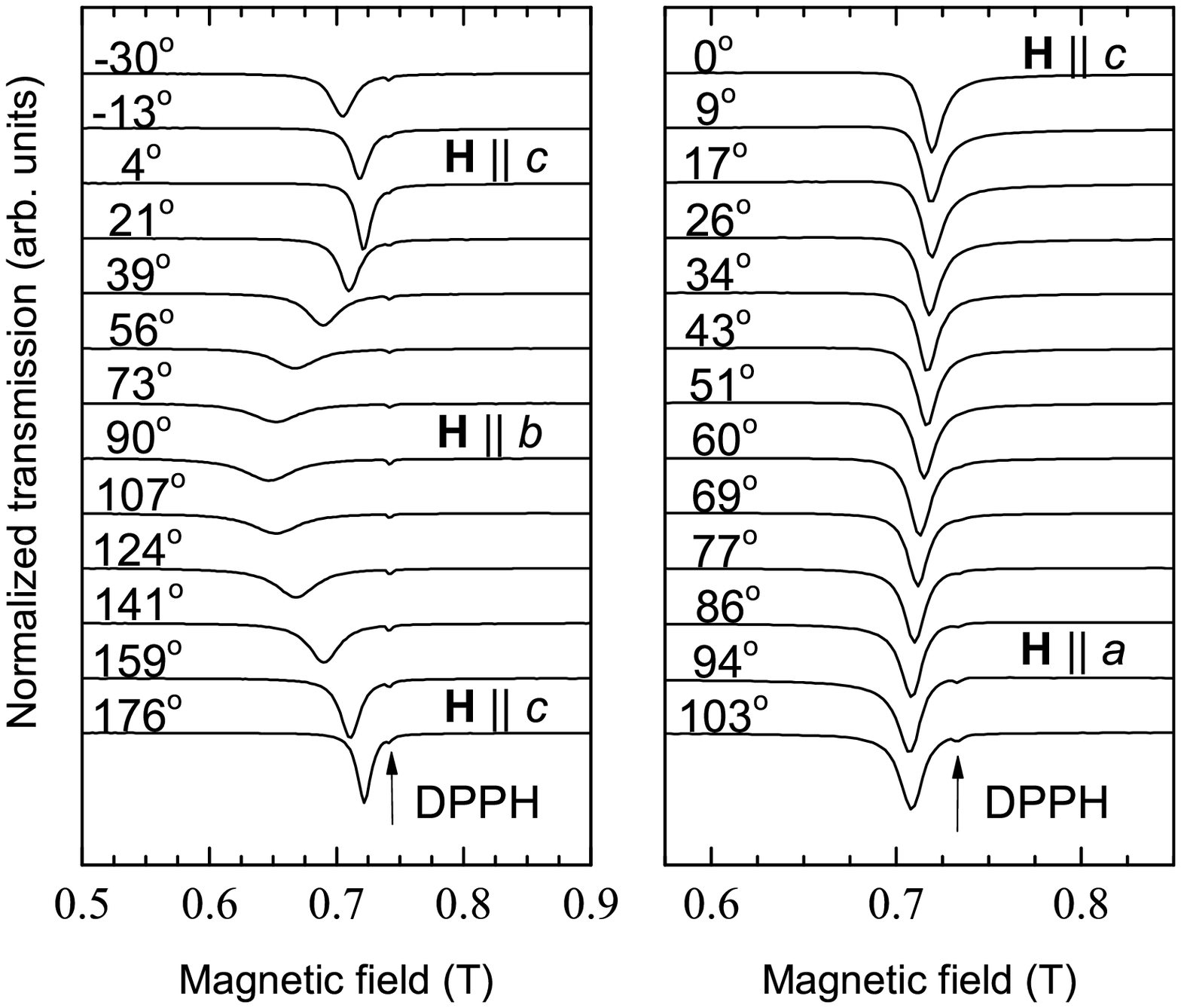}
 \caption{\label{fig9} Normalized ESR absorption curves at frequency $\nu$~=~20.7~GHz for different orientations of the magnetic field
 at $T = 1.3$~K. Left panel: rotation of the magnetic field within $bc$ plane. Right panel: rotation of the magnetic field within $ac$ plane.}
\end{center}
\end{figure}

\begin{figure}[t]
\begin{center}
\includegraphics[width=0.5\textwidth]{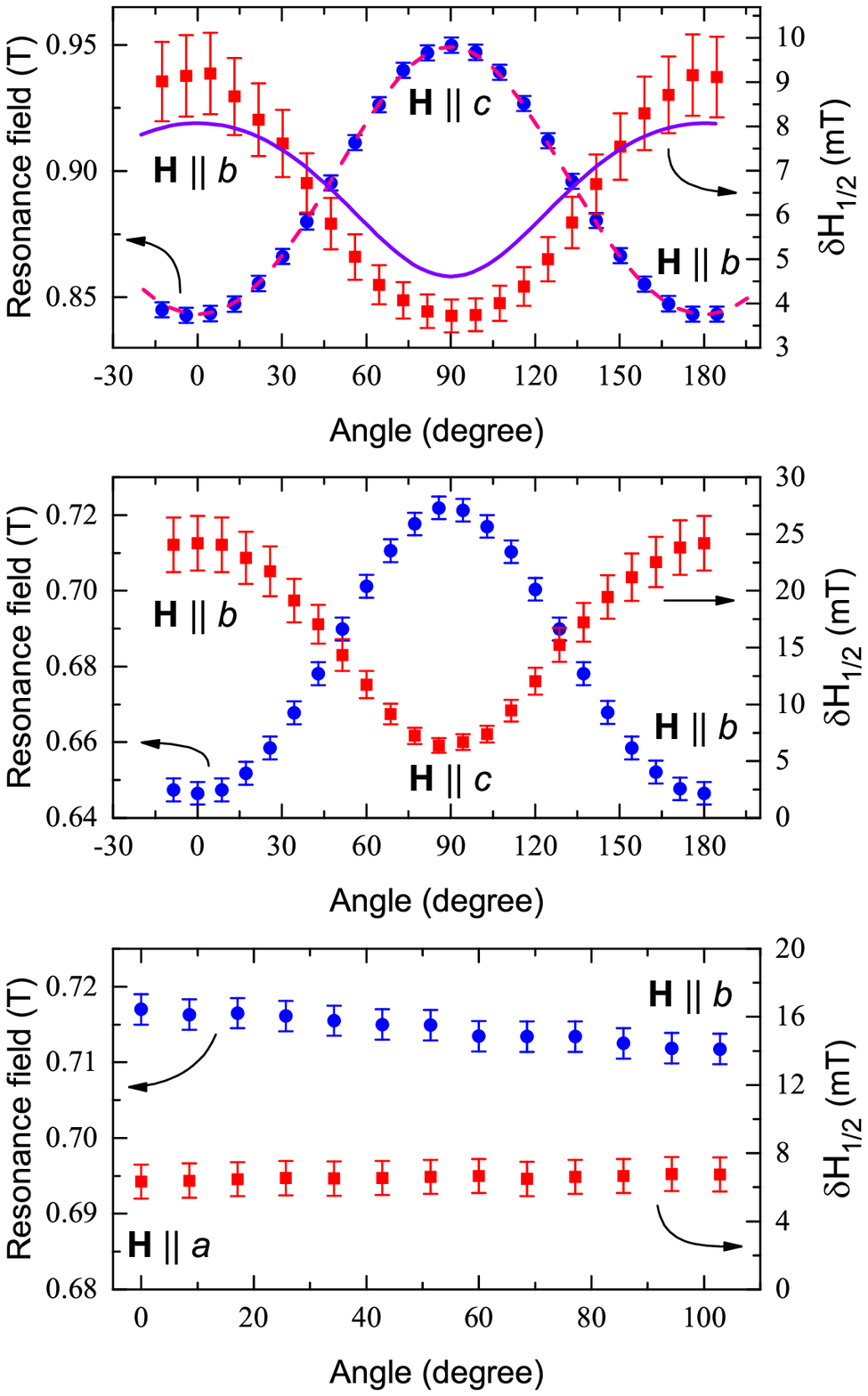}
 \caption{\label{fig10} (Color online) Angular dependencies of the resonance field (circles) and half width at half maximum of ESR
 line $\delta H_{1/2}$ (squares). Upper panel: Angular dependencies for $\nu$~=~27.5~GHz ESR line at $T = 80$~K for field
  in $bc$ plane. Dashed line represents theoretical dependence according to (\ref{Larmor}). Solid line is fit to experimental HWHM using (\ref{HWHM}).
  Middle and lower panels represent dependencies for $\nu$~=~20.7~GHz ESR line taken at $T = 1.3$~K
 for field in $bc$ and $ac$ planes respectively.
 }
\end{center}
\end{figure}

To estimate the value of DM interaction $D$ from the anisotropy of the high-temperature ESR linewidth, we have measured the angular dependencies
of $\delta H_{\frac{1}{2}}$ of 27.5~GHz ESR line rotating the magnetic field in $bc$ plane at temperature 80~K (at which $\delta H_{\frac{1}{2}}$
has  a minimum). The records of ESR lines at different orientations of the magnetic field are given in Fig. \ref{fig9} and the angular
dependencies of the resonance field and linewidth  are given in the upper panel of Fig.\ref{fig10}. Analogous data for the low temperature $T$=1.3
K are presented in the middle and lower panels. High-temperature data demonstrate a significant anisotropy of the linewidth of about 250 percents,
while the anisotropy of the resonance field is about 10 percents and may be  described by the anisotropy of $g$-factor. The low-temperature
angular dependencies for the rotation in the $(bc)$ plane, i.e. from ${\bf H}
\parallel {\bf D}$ to ${\bf H} \perp {\bf D}$ position, demonstrate a remarkable deviation of the linewidth, shown in the middle panel of Fig.
\ref{fig10},
  while for the rotation of the field in
the $(ab)$ plane, when the magnetic field remains perpendicular to DM vector ${\bf D}$, the deviation is within the experimental error, see Fig.
\ref{fig10}, lower panel.

\section{Discussion}

\subsection{Spinon magnetic resonance}\label{spinon res}

According to our observations, in the high temperature range $T > $~3~K, the ESR frequency $\nu$ follows a typical paramagnetic resonance
dependence on the magnitude and orientation of the external field, see, {\it e.g.} Ref. \cite{Abragam}, (equation 1.49):

\begin{eqnarray}
2\pi \hbar \nu = \sqrt{(g_{b} \cos(\theta))^2+(g_{\perp} \sin(\theta))^2} \mu_{B} H \ \ \label{Larmor}
\end{eqnarray}

here $\theta$ is the angle between the external magnetic field and $b$-axis. The corresponding angular dependence of a resonance field  in $bc$
plane is presented in the upper panel of Fig.~\ref{fig10} by a dashed line, there is a good agreement with the experimental resonance fields.

 According to the basic principles of
magnetic resonance, for narrow ESR lines the integral intensity is a measure of the static susceptibility (see, e.g. Ref.\cite{Abragam}, equation
(2.64)). Thus, the ESR intensity should have the same temperature dependence as the  static susceptibility. In Fig. \ref{fig6} we see a good
correspondence of the temperature dependence of ESR integral intensity to Bonner-Fisher calculation \cite{Hatfield} of static magnetic
susceptibility of $S = \frac{1}{2}$ Heisenberg chain with $J = 3.2$~K   for the frequencies, when  $\delta H_{1/2} \lesssim H_{res}$.
 This observation of a typical  susceptibility of 1D $S=1/2$ antiferromagnet  imply, that upon cooling below $T_{CW}$~$\thickapprox$~3~K
  a spin-correlated state like that in
decoupled spin $S=1/2$ chains is formed. This correlated state is also consistent with  the observation of the excitation spectrum in form of a
two-spinon continuum in a neutron scattering experiment. \cite{Povarovetalprivate}
  The comparison of the ESR intensity and Bonner-Fisher susceptibility provides a check
whether the observed  signals are intrinsic for K$_2$CuSO$_4$Cl$_2$ or arise partially from defects or impurities. The good coincidence with the
theoretical susceptibility curve in Fig.~\ref{fig6} confirms   that the experimental ESR intensity  is dominantly intrinsic and not originating
from paramagnetic impurities. Considering the deviation from the Bonner-Fisher susceptibility is within the experimental error we estimate the
concentration of paramagnetic defects as not exceeding  $ \thickapprox 1.5 \%$.

  For an isolated $S = \frac{1}{2}$ Heisenberg chain in an external magnetic field the ESR mode
should appear exactly at a Larmor-type frequency $\nu = g_{\alpha} \mu_{B} H_{\alpha} /(2\pi \hbar)$ as for other spin systems with Heisenberg
exchange.\cite{Al'tshuler, Oshikawa} However, in our ESR experiments we find the shift of ESR mode from Larmor frequency at decreasing the
temperature, as shown in Figs.~\ref{fig2} and ~\ref{fig5}. We believe this shift is due to the influence of uniform DM interaction on excitation
spectrum in K$_2$CuSO$_4$Cl$_2$, in accordance with theory of  Refs.\cite{Gangadharaiah,Povarov,Karimi} This conclusion is confirmed by the
observations of the characteristic anisotropy of the shift of the resonance field, which corresponds to ESR of a spin S=1/2 AFM chain with uniform
DM interaction. Indeed, this anisotropy is consistent with  Eq. (\ref{Freq}), which predicts a negative shift of the ESR field at ${\bf H}\perp
{\bf D}$ and a doublet of lines at ${\bf H}\parallel {\bf D}$. The low-frequency component of doublet should have the positive shift of the field,
thus, the ESR line with a positive shift of the field, observed at ${\bf H}\parallel {\bf D}$, may be interpreted as a lower component of the
doublet.

Consider the ESR frequencies at different orientations of the magnetic field in details. Orientations of magnetic field ${\bf H }\parallel b$ and
${\bf H }\parallel a,c$ correspond to ${\bf H }\parallel {\bf D}$ and ${\bf H }\perp {\bf D}$ respectively as the symmetry of the structure of
K$_2$CuSO$_4$Cl$_2$ implies ${\bf D }\parallel b$. For orientations of the field  ${\bf H }\parallel {\bf D}$ and ${\bf H }\perp {\bf D}$
Eq.~(\ref{Freq}) correspondingly takes the form:
\begin{eqnarray}
 &2&\pi\hbar\nu_{\pm}=\left|g_{b}\mu_{B}H \pm \dfrac{\pi D}{2}\right|,
 \label{freqPAR}
\\
&2&\pi\hbar\nu_{\pm}=\sqrt{(g_{\perp}\mu_{B}H)^{2} +  \left(\dfrac{\pi D}{2}\right)^{2}}. \label{freqPERP}
\end{eqnarray}
According to these expressions, for the magnetic field applied in $ac$ plane single gapped mode with zero-field gap $2 \pi \hbar \Delta = \pi D/2$
should be observed, whereas at ${\bf H }\parallel b$  the ESR doublet should be formed.

ESR spectra obtained in frequency range $\nu >$~2.4~GHz at 0.45~K demonstrate qualitative agreement with theoretical prediction. First, for ${\bf
H }\parallel a$ , ${\bf H }\parallel c$ we observe the ESR mode shifted to lower fields following the scenario of opening of the gap in zero
field. Second, at ${\bf H }\parallel b$ we see the formation of single ESR line shifted to higher fields consisting with the low-frequency
component of the ESR doublet. The upper frequency component of the doublet is not visible. This might be explained by the observations of the
doublet at ${\bf H }\parallel {\bf D}$ in crystals  of K$_2$CuSO$_4$Br$_2$ \cite{Smirnov2} and Cs$_{2}$CuCl$_{4}$ \cite{Smirnov1}, where the
high-frequency component of the  doublet exhibited much lower intensity and disappeared quickly at heating or at the increase of magnetic field
well before the lower component transformed to a Larmor mode. It was empirically shown \cite{Smirnov1,Smirnov2} that the upper limit of the
magnetic field at which the upper component of the spinon doublet may be observed is approximately equal to  $H_{DJ}~=~\sqrt{DJ}/g \mu_{B}$. The
empirical temperature at which doublet components are finally formed and the maximum of resonance field shift for doublet components is achieved
should be lower than
  $T_{DJ}~=~\sqrt{DJ}/k_{B}$.

Fitting the experimental dependence $\Delta \nu$($H$) in the range $\nu >$ 2.4 GHz for ${\bf H }\parallel a$ (see Fig.~\ref{fig5}) with the
relation (\ref{freqPERP}) we get the value of the gap $\Delta = 2.0 \ \pm \ 0.2$~GHz and DM interaction $D = 0.06 \ \pm \ 0.01$~K. Using
Eq.~\ref{freqPAR} we plot the calculated dependence for low-frequency component of doublet expected at  ${\bf H }\parallel b$ in the lower panel
of  Fig.~\ref{fig5}. We observe here the correspondence in the sign of the shift, but the value of the shift reaches only of about one third of
the expected value. Thus, the shift of ESR is lower than expected, and the higher component of the doublet, if it exist, is too weak in intensity
to be detected.

Using the above value of $D = 0.06 \ \pm \ 0.01$~K and the value of $J$=3.2 K, we get $H_{DJ}\simeq$ 0.4~T and $T_{DJ} \simeq$  0.44~K. Thus, the
temperature of our experiment 0.45 K is about $T_{DJ}$ and the  ESR shift and gap opening  should be almost as large as at $T$=0.  This may be
concluded from the comparison with the data on K$_2$CuSO$_4$Br$_2$ presented in Fig.7 of Ref.\cite{Smirnov2}. For this previous investigation the
temperature $T_{DJ}$ was passed during the cooling.  In that case, at $T = T_{DJ}$,  the gap was observed to be  only for 10 percents smaller than
the limit low-temperature value.
 Another indication of that the resonance field at 0.45 K is close to limit value is the observation, that the field ceased changing at cooling to 0.45
K, as presented in Fig.~\ref{fig2}

ESR lines taken at the frequencies below 10 GHz have the resonance fields below $H_{DJ}$=0.4 T. Thus, the condition of our experiments meet the
empirical criterium of the doublet observation in a quasi-1D antiferromagnet with $J^\prime < D$. Therefore, the absence of a good correspondence
of the observed ESR shift to the theory of 1D spin chains with uniform DM interaction and the absence of the upper component of the doublet should
be related to a high value of $J'$, which is of the order of $D$. The formation of the fine structure of the ESR spectrum due to the uniform DM
interaction is predicted for an isolated $S = \frac{1}{2}$ chain, while the interchain exchange comparable with $D$ should naturally disturb the
formation of the fine structure within the spinon continuum. For K$_2$CuSO$_4$Br$_2$  the condition $J' \ll D$ was fulfilled: $J' = 0.03$~K and $D
= 0.27$~K, \cite{Halg, Smirnov2} whereas for the chlorine compound it doesn't take place, while $J' = 0.03 - 0.45$~K and $D = 0.06$~K. Thus, the
energy gap and the spinon doublet are "underdeveloped" because of the action of interchain exchange.

Based on the above concept of the ESR frequencies and linewidth  evolution with temperature and magnetic field, we can treat the anomalous
increase of the linewidth and final transformation of the the Lorentzian resonance line into an  absorption band at frequencies below 2 GHz as a
transition to sub-gap range. Because the value of the interchain exchange is as large as DM energy $D$, (but still much lower than intrachain
exchange $J$) the spinon fine structure is smeared and at a frequency below the gap the spinon-type  ESR may be overlapped with a Larmor-type mode
of the pure Heisenberg system. This overlapping probably results in a band of absorption covering the ranges of the Larmor mode and of the spinon
ESR.

\subsection{High-temperature ESR linewidth}\label{HT linewidth}

We consider the ESR linewidth has a contribution of the spin lattice relaxation processes and of spin-spin interactions. Mainly, isotropic
exchange interaction is known to give no  contribution to the width of the ESR line. The contribution of the symmetric anisotropic exchange
interaction to the ESR linewidth decreases with cooling, whereas an impact of uniform DM interaction increases, resulting in the linewidth rising
with cooling as $1/T$ in the range $T$~$\gg$~$J$. \cite{Fayzullin} The lattice contribution is usually  rising at high temperature. According to
our measurements of the ESR linewidth for K$_2$CuSO$_4$Cl$_2$, shown in Fig.~\ref{fig7}, the linewidth increases with cooling below 80~K
indicating that spin-spin contribution due to uniform DM interaction is dominant. Thus, $\delta H_{\frac{1}{2}}$ measured at 80~K near its minimum
(see lower panel of Fig.~\ref{fig10}) should be mostly due to spin-spin relaxation via uniform DM interaction.  Naturally, at this temperature the
high-temperature approximation $T \gg J/k_{B}$ is valid. We use the results  of the ESR linewidth calculation  by the method of moments applied
for 1D $S = \frac{1}{2}$ Heisenberg chain with uniform DM interaction. In the  high temperature limit ($T \gg J/k_B$), the angular dependence of
$\delta H_{\frac{1}{2}}$ due to spin-spin interaction described by the model Hamiltonian (\ref{Ham}) with DM vector ${\bf D} = (0, \pm D_{b}, 0)$
is as follows: \cite{Fayzullin}

\begin{eqnarray}
\delta H_{1/2}^{DM} = \frac{\pi}{4} \frac{ D^{2}(\theta)}{g(\theta) \mu_{B} J } \ \ \label{HWHM}
\end{eqnarray}
here $g(\theta) = \sqrt{(g_{b} \cos(\theta))^2+(g_{ac} \sin(\theta))^2}$, $D^{2}(\theta) = D_{b}^{2} (1 + (g_{b} \cos(\theta))^2/g^2(\theta))$,
$\theta$ is the same angle  as defined in section \ref{spinon res}. Naturally, the spin chains of both orientations of vector ${\bf D}$ produce
ESR signals with the same linewidth.

Fitting the angular dependence of the linewidth shown in Fig.~\ref{fig10} by Eq.~(\ref{HWHM}) resulted in a solid line in the upper panel of
Fig.~\ref{fig10} This gives the value of $D_{b} = 0.16 \ \pm \ 0.01$~K. Fit shows that the DM interaction describes the angular variation of the
linewidth quite well. The value of $D$ is about three times greater, than the value derived from the spinon resonance spectrum. It may be due to
the  action of the interchain exchange, which makes the shift of the ESR smaller than in abscence of the interchain interaction, causing the
underestimation of $D$ from spinon ESR.

\section{Conclusion}

The observed behavior of ESR in K$_2$CuSO$_4$Cl$_2$ is in a qualitative correspondence with the scenario of spinon ESR formation under the
influence of uniform DM interaction, competing with interchain exchange. The value of DM interaction $D$ was estimated both from the zero-field
gap and from the angular dependence of the ESR linewidth in a high-temperature limit, and is within the interval 0.06~K~$<~D~<$~0.16~K.  The spin
gap mode and the electron spin resonance doublet which should arise under the action of the uniform Dzyaloshinskii-Moriya interaction are
essentially smeared by the interchain exchange which is of the same order of magnitude as the energy of DM interaction. Therefore, we observe a
crossover between the  gapped fine structure of the ESR in 1D antiferromagnetic chain with uniform Dzyaloshinskii-Moriya interaction and a
Larmor-frequency mode of a quasi-1D Heisenberg antiferromagnet.

\section{Acknowledgements}

 We thank Prof. Oleg Starykh for numerous discussions.  We are indebted to Russian Science
Foundation, grant 17-02-01505, supporting ESR investigations, ETHZ team acknowledges support by Swiss National Science Foundation, Division II

\end{document}